\documentclass[manuscript]{aastex}
\usepackage{longtable}

\shorttitle{Effects of Extreme Obliquity Change on Exoplanets}
\shortauthors{Armstrong et al.}

\begin{document}

\title{Effects of Extreme Obliquity Variations on the Habitability of Exoplanets}

\author{J. C. Armstrong}
\affil{Department of Physics, Weber State University, Ogden, UT 84408}
\email{jcarmstrong@weber.edu}

\author{R. Barnes}
\affil{University of Washington, Seattle, WA}

\author{S. Domagal-Goldman}
\affil{NASA Goddard}

\author{J. Breiner, T. R. Quinn, and V. S. Meadows}
\affil{University of Washington, Seattle, WA}

\begin{abstract}
We explore the impact of obliquity variations on planetary habitability in hypothetical systems with high mutual inclination. We show that large amplitude, high frequency obliquity oscillations on Earth-like exoplanets can suppress the ice-albedo feedback, increasing the outer edge of the habitable zone. We restrict our exploration to
hypothetical systems consisting of a solar-mass star, an Earth-mass
planet at 1 AU, and 1 or 2 larger planets. We verify that these systems
are stable for $10^8$ years with N-body simulations, and calculate the
obliquity variations induced by the orbital evolution of the
Earth-mass planet and a torque from the host star. We run a simplified
energy balance model on the terrestrial planet to assess surface
temperature and ice coverage on the planet's surface, and we calculate
differences in the outer edge of the habitable zone for planets with
rapid obliquity variations.  For each hypothetical system, we calculate the outer
edge of habitability for two conditions: 1) the full evolution
of the planetary spin and orbit, and 2) the eccentricity and
obliquity fixed at their average values. We recover previous results
that higher values of fixed obliquity and eccentricity expand the
habitable zone, but also find that obliquity oscillations further
expand habitable orbits in all cases. Terrestrial planets
near the outer edge of the habitable zone may be more likely to
support life in systems that induce rapid obliquity oscillations as
opposed to fixed-spin planets. Such planets may be the easiest to directly characterize with space-borne telescopes.
\end{abstract}


\section{Introduction}

The habitability of a world depends on a host of properties, from observable quantities like its mass and distance from the parent star to those that are difficult if not impossible to measure: atmospheric composition, surface reflectivity, ice, water distribution, etc. In the case of stars as massive as our Sun, detecting  Earth-mass planets in any orbit is difficult with modern technology. In the last decade, attention has turned primarily to the discovery of rocky planets orbiting in the habitable zone (HZ), a shell around a luminous object in which an Earth-like planet could support liquid water on its surface \citep{Dole64,Kasting93,2013ApJ...765..131K}, as these worlds are best-suited for the development and sustainment of life as we know it. In its latest revision \citep{2013ApJ...765..131K}, the HZ is calculated for a highly idealized case, in which many properties of the star, planet, and planetary system are ignored. Following the identification of possible processes that can impact habitability \citep{SuperHab2014}, we explore how gravitational perturbations from additional planets can affect the climate. We find that in many cases, these perturbations can extend the outer edge of the HZ, thereby increasing the number of planets in the galaxy that are potentially habitable.

While the vast majority of work on habitability has used a replica of
the Earth to determine orbits that are potentially habitable, there
are notable exceptions. Some studies explored the habitability of synchronously
rotating planets \citep{Joshi97,Joshi03,Pierrehumbert11,Edson11,Wordsworth11,
  2013ApJ...771L..45Y} . Others \citep{Abe11,Zsom13} considered planets that
are much drier than the Earth. Several studies
\citep{WilliamsPollard02,WilliamsPollard03,Spiegel10,Dressing10}
varied the eccentricity and obliquity of an Earth-like planet and
found that larger values tend to increase the globally averaged
temperature on a planet, while holding the semi-major axis constant.

While these studies made great strides in understanding the Earth's climate's sensitivity to rotation rate, obliquity, and eccentricity, aside from \citet{Spiegel10}, they largely ignored that the latter two properties evolve with time due to gravitational perturbations from other bodies. The Earth maintains a relatively constant axial tilt either due to the presence of the Moon, as suggested by \citet{1993Natur.361..615L}, or due to the inherent stability of Earth's axis, as indicated by \citet{2012Icar..217...77L}. However, as illustrated by \citet{1998PhDT........23W} and  \citet{1998pslv.conf..415W}, changes in the  architecture of our solar system - such as moving Jupiter inwards - can result in dramatic variations in the obliquity of the Earth even with the presence of a Moon.
Still, It has been suggested that the relatively small variations in Earth's obliquity result in a stable climate conducive to the development of life. Adding to this stability is the fact that the orbital eccentricity remains smaller than about 0.05 due to the approximately circular orbits of the large planets of the Solar System. 

It is possible small obliquities and circular orbits are not a requirement for habitability. \citet{WilliamsPollard03} used General Circulation Models (GCMs) to determine how different obliquity variations affect the Earth's climate. They found that Earth-like planets with high obliquities were no more likely to experience extreme runaway greenhouse or snowball Earth events, making them just as habitable as Earth. Later \cite{Spiegel10} examined how large eccentricity oscillations affect the climates of rocky exoplanets. They found that in some cases, planets could break out of a snowball event during periods of high eccentricity. It is the goal of the current study to build on these previous results and explore self-consistent models of the climates of planets that experience rapid, large amplitude, and possibly chaotic oscillations of eccentricity and obliquity.

Orbit-induced seasonal effects like the ice-albedo feedback determine the limit of the outer edge of the HZ. As the surface temperature drops, volatile ices such as CO$_2$ and water can condense on the surface. The high albedoes of their solid phases inhibit a planet's ability to absorb solar radiation reducing the temperature further. Mars, if it possessed sufficient surface water, would have been in danger of falling prey to these snowball episodes. Geological evidence exists for these episodes in Earth's past \citep{1998Sci...281.1342H}. For the Earth, a dynamic CO$_2$ recycling system works to offset the negative effects of these events on timescales of millions of years \citep{Walker81}. As the planet cools, weathering rates slow down and lock CO$_2$ in the atmosphere. As the CO$_2$ builds up, the greenhouse effect increases, eventually melting the ice. On early Mars, as now, such robust CO$_2$ cycling could not have been enough to resurrect the planet from these snowball events. From seasonally resolved modeling, it seems that other factors, in particular orbital and obliquity variations, would have to play a role \citep{Paper1}. Mars' obliquity has probably undergone significant evolution in the past \citep{1993Natur.361..608L} due to gravitational torques by the other planets in the Solar System, a property which may have permitted at least episodic liquid water at the surface.

With this solar system context in mind, we turn our attention to
predicting the habitability of exoplanets. We do not anticipate the
photometric and spectroscopic data needed to characterize extrasolar
planets until the launch of the James Webb Space Telescope in $\sim
2017$. Spacecraft like Kepler and ground based radial velocity surveys
are already returning data that can pin down, through computational
analysis, important orbital parameters that impact the climate such as eccentricity, timing of
perihelion passage, and the evolution of the spin axis of the planet. Through the coupling of these data to N-body
simulations and simple, fast climate codes, a more comprehensive
picture of the HZ of a system can be obtained. This type of analysis
can be used to  prioritize future characterization observations, which
are likely to be challenging, expensive, and based on precious little
information. In this study we find that planetary system architecture
--- that is, the distribution of masses and orbits of the other
planets in the system --- can play a significant role in defining the extent of the habitable zone.

The rotational evolution of the bodies in our Solar System can be accurately modeled because the masses and orbits of the planets are extremely well measured. For exoplanets, the situation is more difficult. Radial velocity surveys \citep[e.g.][]{Butler06} are only able to place a lower bound on mass, and cannot measure the relative inclination between orbital planes. Kepler can constrain inclinations \citep{Fabrycky12}, but is heavily biased toward the discovery of planets in coplanar configurations. While these systems are analogous to our Solar System, we note that the one system for which astrometry (which is not biased toward any particular inclination) has measured a mutual inclination, $\upsilon$ Andromedae \citep{2010ApJ...715.1203M}, the relative inclination is 30$^\circ$. Moreover, studies that predict the large eccentricities of exoplanet orbits, simultaneously predict large inclinations \citep{MarzariWiedenschilling02,Chatterjee08,Raymond10,Barnes11}. It is possible, perhaps likely, that there exists a population of planetary systems with large inclinations and a potentially habitable planet. These architectures will induce much larger changes in orbital inclination, which in turn induces large obliquity oscillations. The GAIA mission may be able to determine the range of architectures for giant planets \citep{Casertano08,Sozzetti13}.

As no rocky planet is currently known to orbit with sibling planets with high mutual inclinations, we explore the phenomena with 17 hypothetical, dynamically stable systems. We find that there is a direct link between the orbital architecture of a planetary system and the possible range of climate conditions on a potentiality habitable planet. We use these models to constrain the orbital conditions of a hypothetical planet and find that orbital and rotational evolution tend to push the outer edge of the HZ out, relative to planets where no evolution occurs.

Below, we outline a model that links physically realistic orbital architectures to the spin evolution of a hypothetical Earth-like planet, and finally to its climate. In Section \ref{sec:orbit}, we discuss the motivation behind the systems we have modeled in an effort to obtain a set that spans a range of orbital elements. In Section \ref{sec:obl}, we outline the model used to evolve the spin axis of the planet. In Section \ref{sec:climate}, we present a simplified energy balance model designed to be robust across wide variations of these orbital changes and fast enough that million-year integrations require only second of computational time. We then present the results of these models in Section \ref{sec:results} followed by a discussion in Section \ref{sec:discussion}. 

\section{Orbital simulations}
\label{sec:orbit}

In order to make a first assessment of the potential of similar
architectures to support habitable worlds, we created 17 hypothetical
systems with moderate inclinations, always initially including an
Earth-like planet on a circular orbit 1 AU from a solar-mass star. The
orbital architectures are arbitrary, but consistent with the
distribution of orbital elements of known planets.
While the potentially habitable world is always the same, its siblings have a wide array of properties. Systems consist of 2--3 planets, with eccentricities ranging from 0--0.3 and mutual inclinations from 10--30$^\circ$. The orbital properties of these cases, in astrocentric coordinates, are presented in Table \ref{tab:systemsummary}.

These systems were selected after careful consideration of orbital stability. We numerically integrated each case for 100~Myr in ``hybrid'' mode with {\em Mercury} \citep{1999MNRAS.304..793C} and confirmed that the evolution of every orbital element appeared periodic. Additionally, energy was conserved to better than 1 part in $10^6$, which is sufficient for numerical accuracy \citep{2004ApJ...611..494B}.

We then rotated each system such that the reference plane corresponds to the fundamental plane. For those systems with super-Jupiter planets, this conversion can change some orbital elements significantly. We then reran each case at very high resolution for $\sim$ 1 Myr, conserving orbital angular momentum to 1 part in $10^{12}$. We are therefore confident that no numerical inaccuracies are propagated into the rotational calculations.

\section{Obliquity modeling}
\label{sec:obl}

\begin{figure}
\includegraphics[scale=0.5]{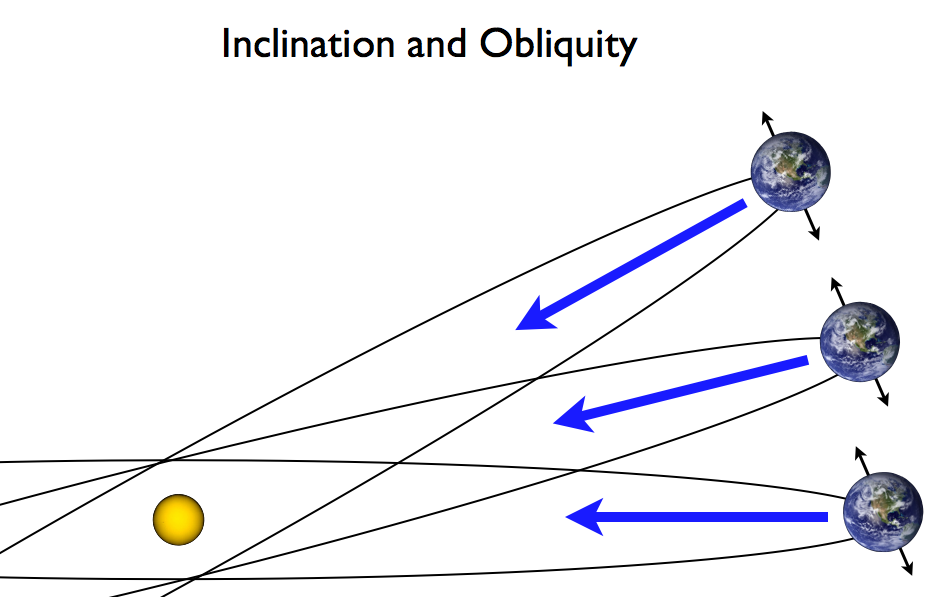}
\caption{A simplified schematic illustrating how the evolution of the inclination leads to an evolution in obliquity. As the inclination of the orbit increases, the spin axis continues to point at a fixed position in space, causing the angle between the spin axis and the fundamental plane to increase. In this figure, the inclination changes in such a way that the obliquity goes from a starting value of 23.5 degrees to 0 degrees.}
\label{fig:schematic}
\end{figure}

Using the results from the orbital runs described above, we employ the obliquity model of \citet{1986A&A...157...59L} as used in previous orbit coupled modeling by \citet{Paper1}. There are two primary factors that influence the evolution of the obliquity.  First, variations in the geometry of the orbit governed by the overall system architecture are present in the orbital elements derived from the N-body simulations. The gravitational influence of the other massive bodies in the system affect the eccentricity, $e$, the inclination, $i$, the argument of perihelion, $\omega$, and the longitude of the acceding node, $\Omega$, of the Earth-sized planet, here cast in terms of the eccentricity-inclination variables 

\begin{eqnarray}
	h & = & e \ sin\left(\varpi\right), \\
	k & = & e \ cos\left(\varpi\right), \\ 
	p & = & sin\left(\frac{i}{2}\right)sin\left(\Omega\right), \\
	q & = & sin\left(\frac{i}{2}\right)cos\left(\Omega\right).,
\end{eqnarray}

\noindent  where $\varpi = \Omega + \omega$ is longitude of perihelion. Changes in parameters like the inclination result in changes in the spin axis orientation relative to the fundamental plane, i.e. the obliquity, as illustrated in Figure \ref{fig:schematic}. In other words, the rotational angular momentum is decoupled from the orbital angular momentum.

In addition to these geometric factors, the direct torques from the central star are included as a term, $R\left(\psi\right)$, in the precession, $p_A$, and obliquity, $\psi$, evolution equations,

\begin{eqnarray}
	\frac{dp_A}{dt} & = & R\left(\psi\right) - cot\left(\psi\right)\left[A(p,q)sin(p_A) + B(p,q) cos(p_A)\right] - 2 C(p,q) - p_g, \\
	\frac{d\psi}{dt} & = & -B(p,q) sin(p_A) + A(p,q) cos(p_A).
\end{eqnarray}

\noindent with,

\begin{eqnarray}
	R\left(\psi \right) & = & \frac{3 k^2 M_{\*}}{a^3 \nu}E_D S_0cos\left(\psi\right), \\
	A(p,q) & = & \frac{2}{\sqrt{1 - p^2 - q^2}} \left[ \dot{q} - pC(p,q)]\right] ,\\
	B(p,q) & = & \frac{2}{\sqrt{1 - p^2 - q^2}} \left[ \dot{p} - qC(p,q)]\right] \\
	C(p,q) & = & q\dot{p} - p\dot{q}, \\
	S_0 & = & \frac{1}{2}\left(1 - e^2\right)^{-3/2} - 0.522\times10^{-6}.
\end{eqnarray}

\noindent Here, the semi-major axis, $a$, is measured in AU, the planet's angular velocity, $\nu$, is measured in $rad \ day^{-1}$, $M_{\*} = 1.0$ in solar units, $k^2 = G M_{\*}/4 \pi^2$ with $G$, the gravitational constant, measured in units of $AU^3 M_{\odot}^{-1} day^2$, and $E_D$ is the dynamical ellipticity, a measure of the non-sphericity of the planet.  Finally, the relativistic precession is accounted for with

\begin{equation}
	p_g = \frac{\kappa_r}{2\left(1 - e^2\right)}
\end{equation}

\noindent where the value of $\kappa_r$ is 

\begin{equation} 
	\kappa_r = \frac{n^3 a^2}{c^2\left(1 + M_p/M_{\*}\right)},
\end{equation} 

\noindent where $c$ is the speed of light, $M_p$ is the planet's mass in units of solar masses, and $n$ and $a$ are related via Kepler's law

\begin{equation}
	n^2 a^3 = k^2 \left(1 + M_p/M_{\*}\right)
\end{equation}

\noindent as outlined in \citet{1986A&A...164..437L, 1986A&A...157...59L}. The values used for these model parameters are listed in Table \ref{tab:oblparams}.

For each of the 17 orbital runs, the Earth-mass planet in the simulation has a 1 day rotation with an Earth-like dynamical ellipticity and orbits a Sun-like star. The obliquity, precession angle, and rates of change in these parameters are computed each time step from the input orbital parameters, and those results are used as the initial conditions for the following time step using a fourth-order Runge-Kutta integrator. These results are then available for further coupling to the energy balance model described below. The primary limitation to the method is that only direct torques from the central body are included. Also, there is an implicit assumption that the planet is rapidly rotating and that it can be accurately described by the dynamical ellipticity relative to the central body. Since none of our models involves tidally locked worlds or close passing planets or moons, these assumptions are justified.

\begin{figure}
\includegraphics[scale=0.6]{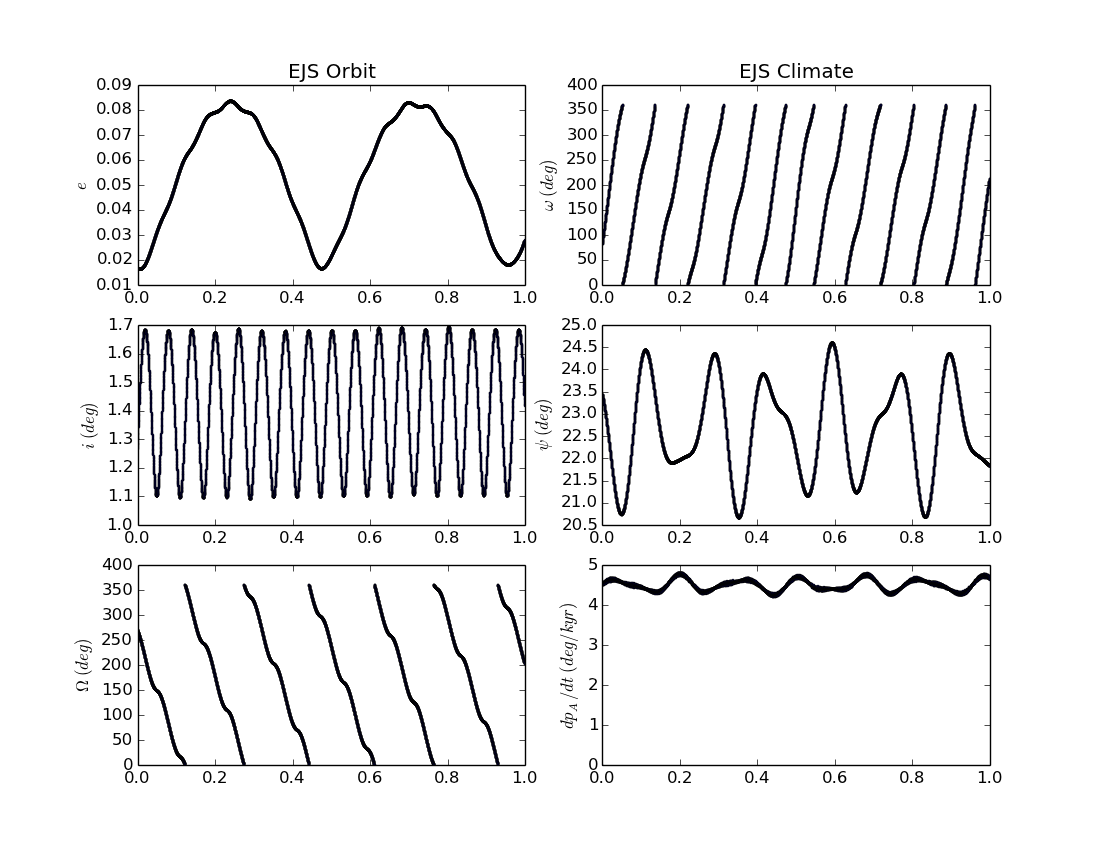}
\caption{Orbital-rotational  results for System 1, Baseline, the Earth-like comparison system. The left column shows the variations of eccentricity, inclination, and longitude of ascending node, and the right column shows the variations of the argument of perihelion, obliquity, and precession rates.}
\label{fig:EJS}
\end{figure}

\begin{figure}
\includegraphics[scale=0.6]{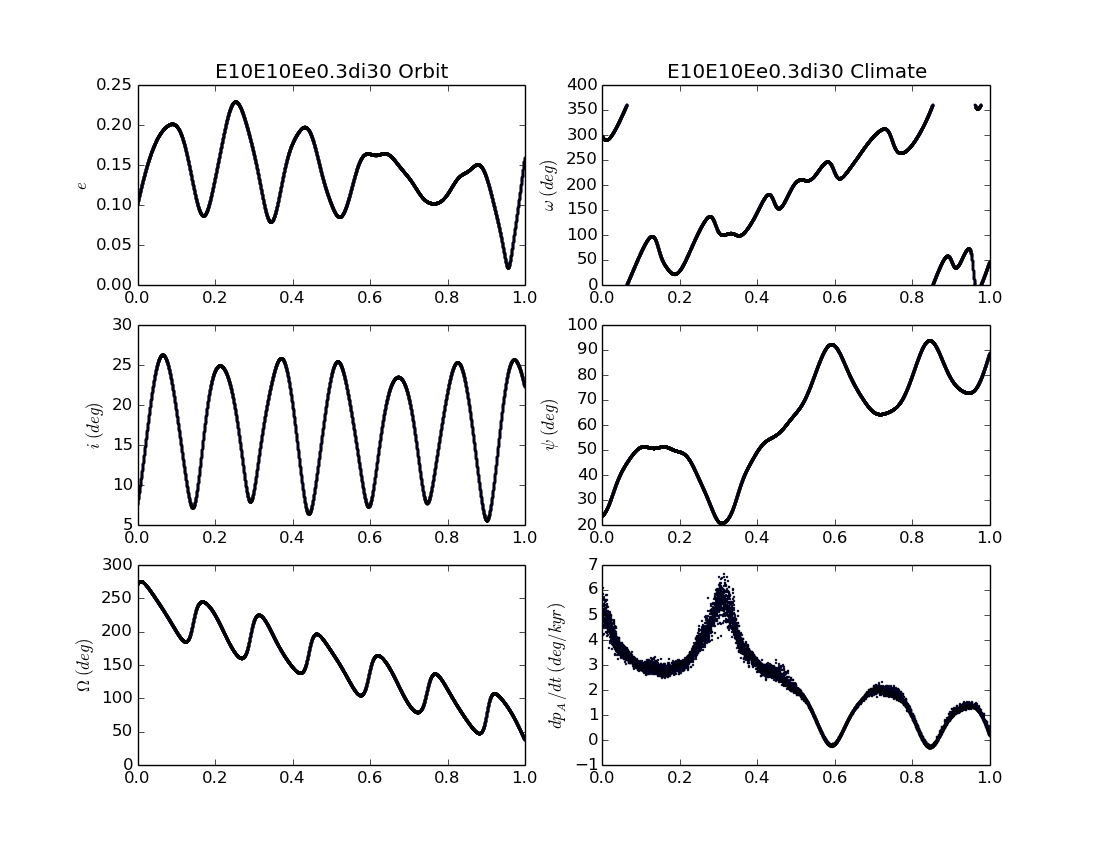}
\caption{Orbital-rotational  results for System 2. The left column shows the variations of eccentricity, inclination, and longitude of ascending node, and the right column shows the variations of the argument of perihelion, obliquity, and precession rates.}
\label{fig:E10E10Ee0.3di30}
\end{figure}

\begin{figure}
\includegraphics[scale=0.6]{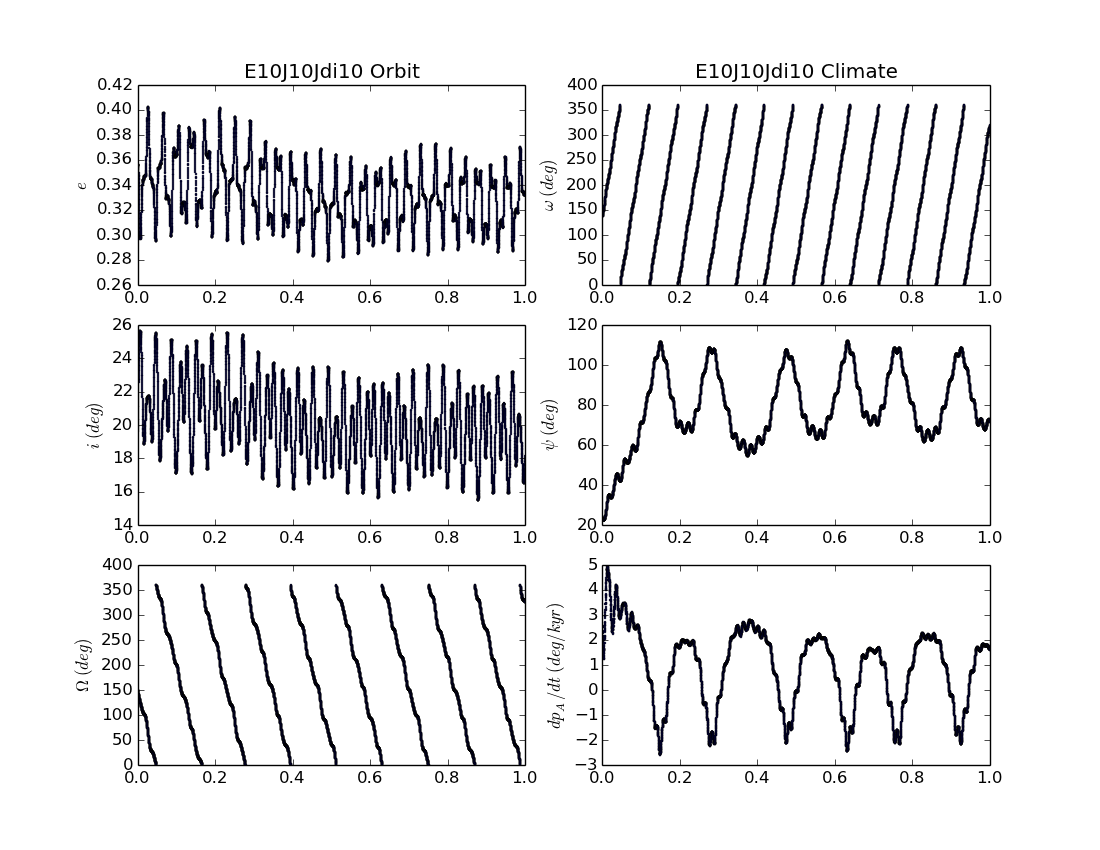}
\caption{Orbital-rotational  results for System 3. The left column shows the variations of eccentricity, inclination, and longitude of ascending node, and the right column shows the variations of the argument of perihelion, obliquity, and precession rates.}
\label{fig:E10J10Jdi10}
\end{figure}

\begin{figure}
\includegraphics[scale=0.6]{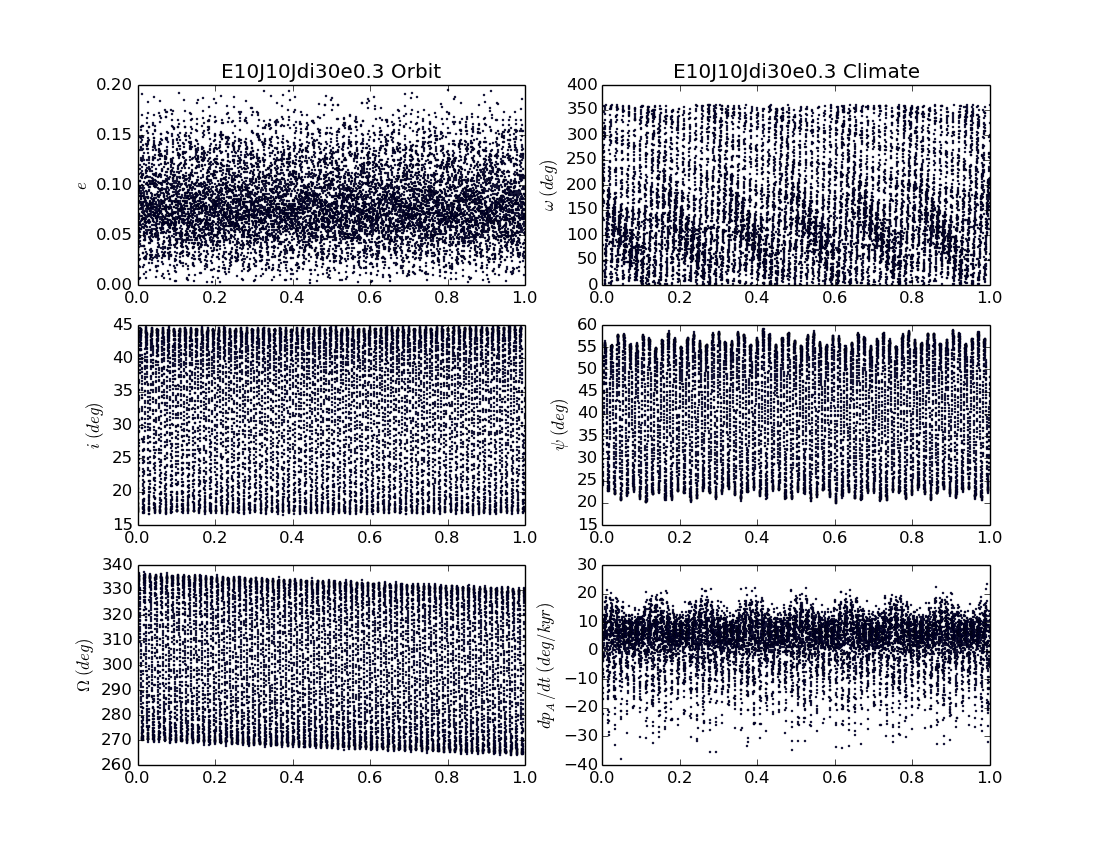}
\caption{Orbital-rotational  results for System 4. The left column shows the variations of eccentricity, inclination, and longitude of ascending node, and the right column shows the variations of the argument of perihelion, obliquity, and precession rates.}
\label{fig:E10J10Jdi30e0.3}
\end{figure}

\begin{figure}
\includegraphics[scale=0.6]{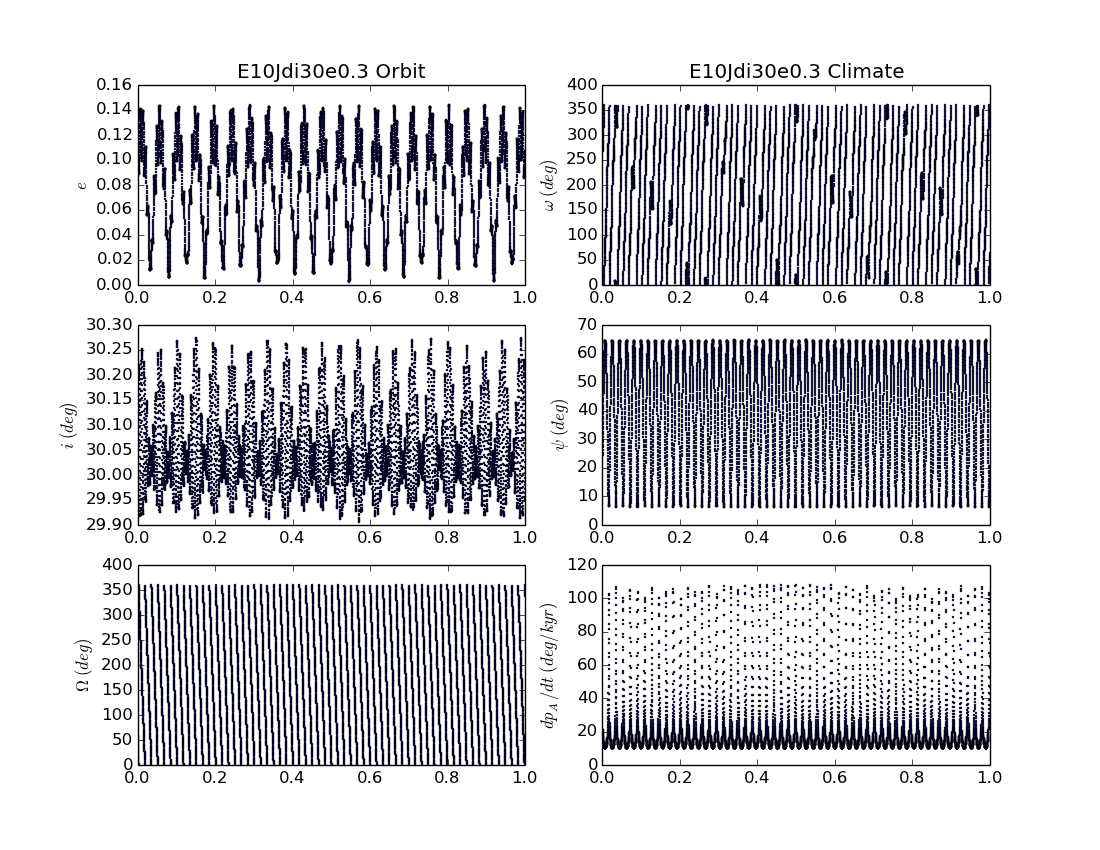}
\caption{Orbital-rotational  results for System 5. The left column shows the variations of eccentricity, inclination, and longitude of ascending node, and the right column shows the variations of the argument of perihelion, obliquity, and precession rates.}
\label{fig:E10Jdi30e0.3}
\end{figure}

\begin{figure}
\includegraphics[scale=0.6]{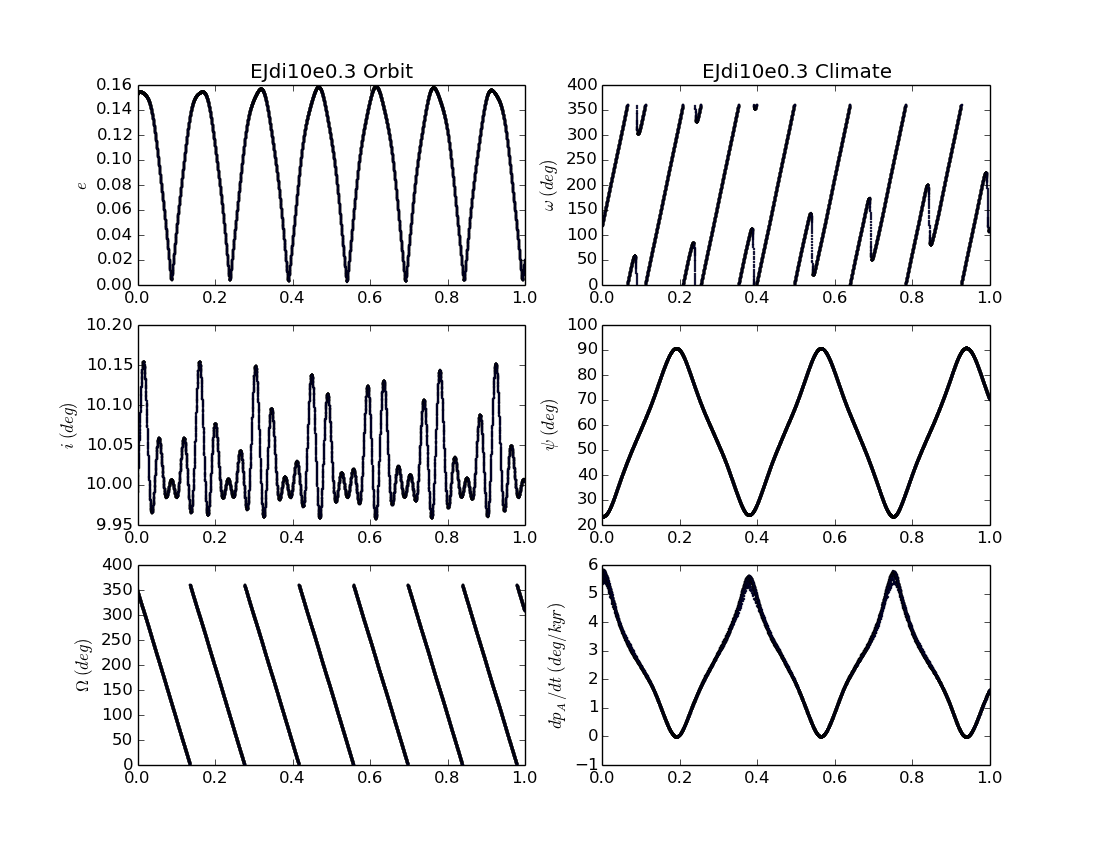}
\caption{Orbital-rotational  results for System 6. The left column shows the variations of eccentricity, inclination, and longitude of ascending node, and the right column shows the variations of the argument of perihelion, obliquity, and precession rates.}
\label{fig:EJdi10e0.3}
\end{figure}

\begin{figure}
\includegraphics[scale=0.6]{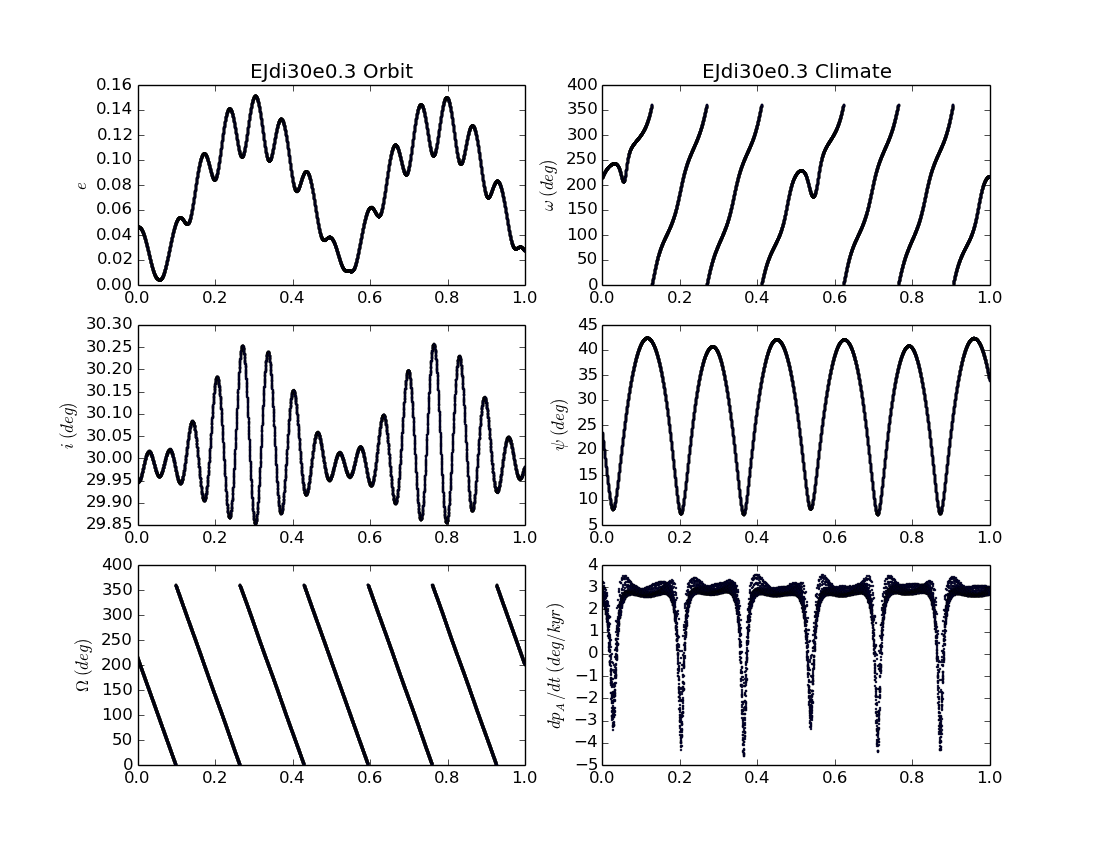}
\caption{Orbital-rotational  results for System 7. The left column shows the variations of eccentricity, inclination, and longitude of ascending node, and the right column shows the variations of the argument of perihelion, obliquity, and precession rates.}
\label{fig:EJdi30e0.3}
\end{figure}

Of the 17 systems studied, we selected seven to analyze that represented the full spectrum of outcomes from the models:

\begin{enumerate}
	\item Earth-Jupiter-Saturn system for comparison (System 1, Baseline - Figure \ref{fig:EJS}).
	\item Two systems with high mean obliquity but modest variations (Systems 2 and 3 - Figures \ref{fig:E10E10Ee0.3di30} and \ref{fig:E10J10Jdi10}).
	\item Two systems with wide and rapid variations in obliquity (Systems 4 and 5 - Figures \ref{fig:E10J10Jdi30e0.3} and \ref{fig:E10Jdi30e0.3}).
	\item Two systems with wide and slow variations in obliquity (Systems 5 and 6 - Figures \ref{fig:EJdi10e0.3} and \ref{fig:EJdi30e0.3}).
\end{enumerate}

Figure \ref{fig:EJS} illustrates the baseline run with the (moonless) Earth, Jupiter, and Saturn. The panels on the left represent the relevant orbital parameters used in the obliquity calculations, and the right hand panels are the parameters of interest for the climate calculations.  These solutions are qualitatively similar to recent work (e.q. \citet{2012Icar..217...77L}) but differ in the detailed periodicities and magnitudes of the obliquity and precession due to the specific orbital solution used in the model. However, the climatically important parameters have relatively low amplitude and are slowly varying compared to other models.

Figure \ref{fig:E10E10Ee0.3di30} shows a model with the Earth-mass planet in a system with two other planets 10 times its size, with large eccentricity variations and high mutual inclination. This architecture results in wide swings in obliquity from the starting value to about 80 degrees, but has relatively slow variations. 

Figure \ref{fig:E10J10Jdi10} shows a model with two 10 Jupiter-mass planets orbiting in the system with the Earth-mass planet, again with high eccentricity, high mutual inclinations, and an obliquity that oscillates around 85 degrees with a period of approximately 200,000 years.

Figures \ref{fig:E10J10Jdi30e0.3} and \ref{fig:E10Jdi30e0.3} illustrate some of the cases where the obliquity variations become rapid (periods of $<$ 16,000 years) at relatively high amplitudes between 10 and 60 degrees. In each of these cases, there is high mutual inclination and large eccentricities due to two 10 Jupiter-mass planets in Figure 4 and one Jupiter-mass planet in Figure 5.

Lastly, Figures \ref{fig:EJdi10e0.3} and \ref{fig:EJdi30e0.3} show an Earth-like planet that has a Jupiter mass planet in the system with a mutual inclination of 10 degrees (Figure  \ref{fig:EJdi10e0.3}) and 30 degrees (Figure  \ref{fig:EJdi30e0.3}) resulting in a wide range in obliquity that varies more slowly than the previous cases.

The precession rates illustrated in the figures are positive, as expected, except near $\psi = 0$ or $\psi > $ 90 degrees. These are numerical artifacts of the geometry of the semi-analytical model. As the obliquity goes beyond 90 degrees, the precession rate reverses due to the fact the``south'' pole is now the ``north'' pole. Additionally, the cotangent term in Equation 5 makes the precession poorly defined near 0 degrees. From a climate perspective, the precession is unimportant at low obliquities, and this is only relevant for ``bookkeeping''.

\section{A simplified energy balance model}
\label{sec:climate}

With the orbital variations and obliquity calculations in hand, we assess the surface conditions using a simplified energy balance model (EBM). This  calculation requires a model that can simulate planets on the timescales for glacier growth/retreat: $10^4$ - $10^5$ years. GCM's are generally too complex to be practical on these timescales. And both GCM's and detailed EBM's, to a greater or lesser extant, require detailed information about the planet or assumptions based on relevant observations. In the case of exoplanets, especially those with widely varying orbital parameters, making these assumptions is difficult if not impossible. To include enough detailed physics to allow for direct calculation of climate effects in absence of observation would render our code too cumbersome.  To act as a first order comparison between the other systems and the baseline model, with as few input parameters as possible, we have developed a simplified energy balance model to examine the conditions on the surface. While the lack of detail makes specific climate predictions difficult for individual planets, general comparisons should be valid.

We model the surface as a 1-D latitude grid of 90 bands from -90 S to +90 N and the atmosphere as a single slab grey absorbing layer. We modify a simplified model outlined by \citet{Paper1} to include basic deposition and evaporation of water ice over the seasonal cycle. 

For each time step in the million-year orbital-obliquity model, we model three years of the seasonally resolved climate, one year to act as a spin-up from the initial conditions, and two to compute global surface properties. Each model year starts with $t = 0$ at the time of perihelion passage, $T_{peri} = 0$. From this, we can compute the eccentric anomaly, $E$, from the implicit equation

\begin{equation}
	T_{peri} = \left(\frac{a^3}{G M_{*}}\right)^{1/2}\left(E - e sinE\right),
\end{equation}

\noindent using Newton's method (where all of the values above are measured in MKS units). The eccentric anomaly is related to the true anomaly, $f$, by

\begin{equation}
	f = 2 \arctan \left[\left(\frac{1 + e}{1 - e}\right)^{1/2} tan\left(\frac{E}{2}\right)\right].
\end{equation}

\noindent With the true anomaly, we compute the instantaneous distance to the star,

\begin{equation}
	r = \frac{a\left(1 - e^2\right)}{1 + e\cos f}.
\end{equation}	

\noindent The instantaneous stellar distance allows us to compute the incident stellar radiation,

\begin{equation}
	S_p = \frac{L_*}{4\pi r^2},
\end{equation}

\noindent where $L_*$ is the stellar luminosity. From this, we can compute the daily mean top-of-atmosphere instellation at any point on the globe,

\begin{equation}
	I_d = \frac{S_p}{\pi}\left[\eta \sin\delta_{*}\sin\delta + \sin(\eta)\cos\delta_*\cos\delta\right],
\end{equation}  

\noindent where $\eta$ is the half-angle of daylight --- a measure of
the length of the day --- at a given latitude $\delta$ and the substellar latitude, $\delta_* = \psi \sin(L_s)$, is determined by the obliquity, $\psi$, and the stellar longitude, $L_s = L_{sp} + f$ where $L_s = 0$ is the northern hemisphere vernal equinox. $L_{sp} = L_{sp0} - p_A - \omega$ is the solar longitude of perihelion, determined by the arbitrary constant $L_{sp0}$ (in our case, 90 degree represents northern hemisphere summer solstice), the spin precession, $p_A$, and argument of perihelion, $\omega$. The half-angle of daylight, $\eta$ is given by

\begin{eqnarray}
	\cos \eta   =  - \tan \delta \tan \delta_*  & &\mid \delta \mid < 90 - \mid \delta_* \mid,  \\
	\cos \eta  =   -1 \ \ \ \ \ \ \ \ \ \ \ \ \ & & \delta  \leq -90 + \delta_*  \ or \   \delta  \leq 90 + \delta_* \\
	\cos \eta   =  1 \ \ \ \ \ \ \ \ \ \ \ \ \ \ \  & &  \delta  \geq 90 - \delta_*  \ or \   \delta  \leq 90 - \delta_*  
\end{eqnarray}

To estimate the surface temperature, we first determine the energy balance at the surface at each latitude bin,

\begin{equation}
	\Delta E = I_d (1 - A) - \epsilon_s \sigma T_e^4 + F_{surf},
\end{equation}

\noindent where $A$ is the surface albedo of either ground or ice depending on local conditions, $\epsilon_s$ is the surface emissivity, $T_e$ is the planet's atmosphere-free equilibrium temperature, and $F_{surf}$ is the heat flux from the surface. The atmosphere is modeled as a single slab with opacity $\tau$ that is equivalent to the number of absorbing layers required to achieve a surface temperature of $T_s$, related to $T_e$ by

\begin{equation}
	T_s = (1 + \tau)^{1/4} T_e
\end{equation}

\noindent The $\tau$ parameter is essentially a one-dimensional model of the greenhouse effect. That is, adjusting $\tau$ as a free parameter allows us to include a contribution to greenhouse warming, with $\tau = 0.095$ reproducing the Earth-like global surface temperature in our baseline model. To estimate the surface temperature, we set $\Delta E = 0$, solve first for the equilibrium temperature and then for the surface temperature. 

Ice deposition is handled parametrically in the model by choosing a global ice deposition rate that comes into effect when the surface temperatures dip below 273 K. At this point, the surface albedo is set to the value for ice and ice is allowed to accumulate. If the surface temperature exceeds 273 K in the presence of ice, the surface temperature is held at 273 while the deposits are evaporated according to the difference between what the  temperature of the surface would be in the absence of ice and the freezing point of water, $T_{ice} = 273$. In this case, the mass loss rate in units of $kg \ s^{-1} \ m^{-2}$ is given by

\begin{equation}
	\frac{\Delta M_{ice}}{\Delta t} = \frac{\sigma\left(T_{ice}^4 - T_{s}^4\right)}{L_h}
\end{equation}

\noindent where $L_h = 3.34\times10^5$ is the latent heat of fusion for ice in $J \ kg^{-1}$. Once the ice is gone, the albedo is reset to the surface value and the surface temperature evolves normally.  The ice deposition rate is set such that the thickness and extent of the ice-snow regions are roughly comparable with values for present-day Earth in the baseline model.

In summary, the only ``parameters'' are the albedo of land/ice, the
atmospheric opacity (which is varied as a free parameter in the
study), the ice deposition rate, and a parameterized version of
thermal exchange with the surface. To calibrate the model, we chose
reasonable values for the albedo, tune the deposition rate to get
reasonable ice caps, and modify the atmospheric opacity and surface
flux to hit a global mean temperature of $\sim$288 K. The parameters
of the model are outlined in Table \ref{tab:clmparams}. 

This approach has several important features:

\begin{enumerate}
	\item The model is extremely fast, allowing us to run thousands of cases over millions of years to explore the full parameter space of the model.
	\item The physics is extremely simplified and coupled tightly to the orbital parameters. Therefore, we can see the first order effects of the orbital variations, the main goal of this study.
	\item The model is determined by only three free parameters: The opacity (or number of absorbing layers), the surface heat flux, and the ice deposition rate.
	\item The final results are normalized to the baseline run, which, while limiting what we can say about the specific climate, allows for a quantitative comparison of the outer edge of the HZ for the hypothetical planets and the modern Earth.
\end{enumerate}

\begin{figure}
\hspace*{+1.50in}
\center
\includegraphics[scale=0.4]{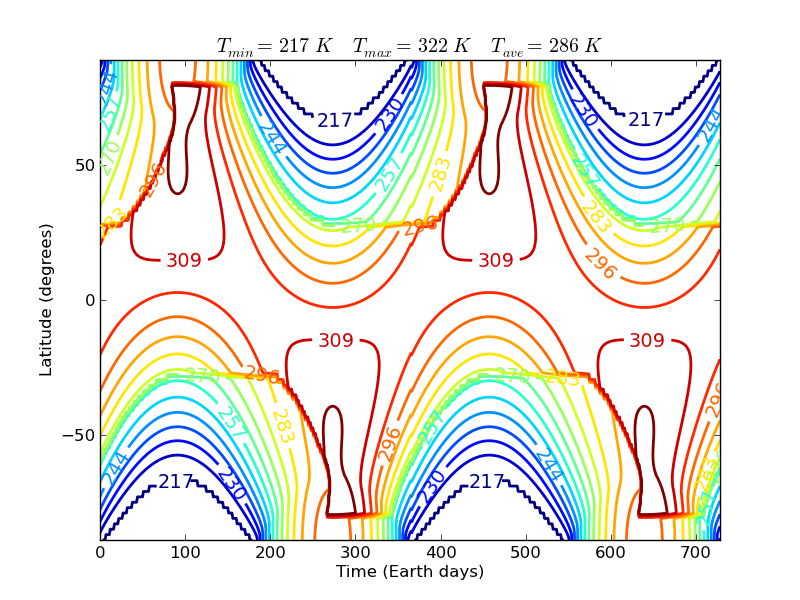} \\
\includegraphics[scale=0.4]{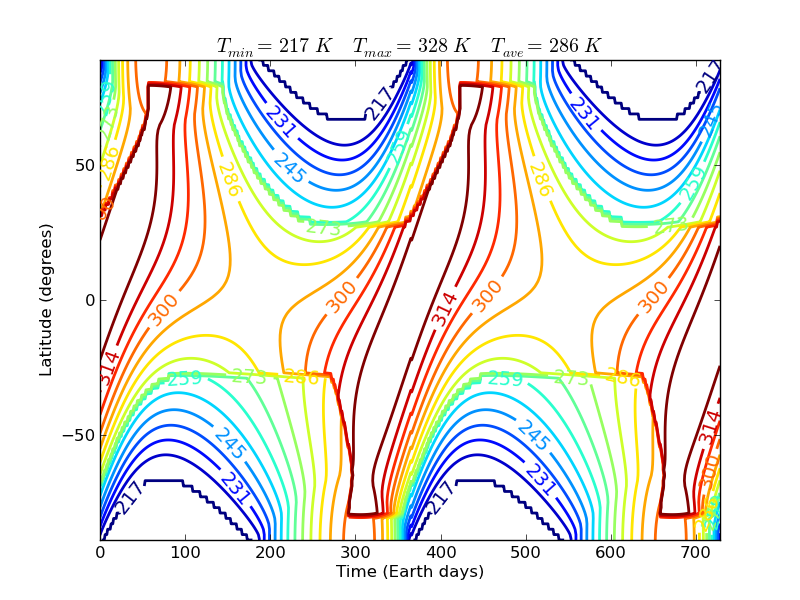} \\
\includegraphics[scale=0.4]{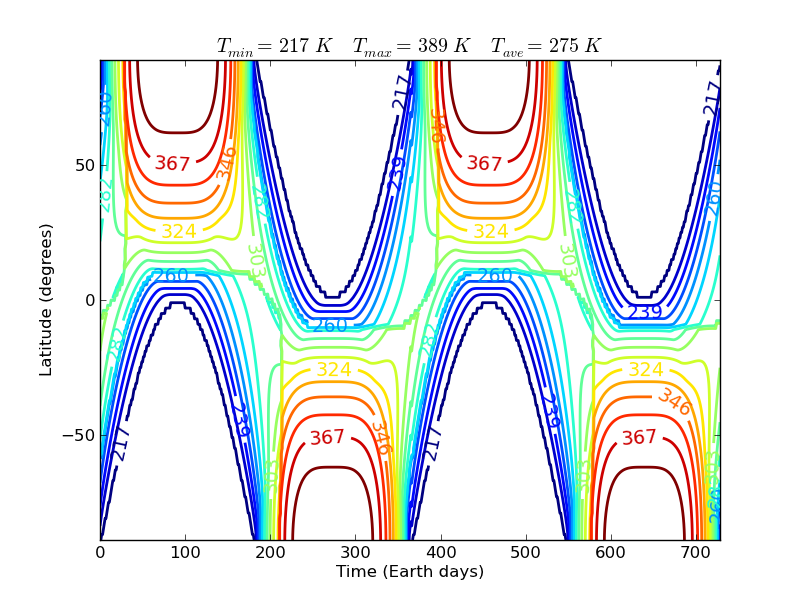} \\
\caption{Baseline climate models for an Earth-like planet with $\psi = 25$, $e = 0.0$ (top), $\psi = 25$, $e = 0.15$ (middle), and $\psi = 90$, $e = 0.0$ (bottom). The discontinuities are caused by the change in albedo between an ice/snow covered and ice-free surface.}
\label{fig:climatebl}
\end{figure}

This model reproduces the first-order climate effects and the ice-albedo feedback (see Figure \ref{fig:climatebl}). Each panel plots the two model years after the spin-up year. The top plot is the baseline run, showing the surface temperatures for our nominal ``Earth''. The discontinuities result from the abrupt change in surface albedo. The second model is identical to the first, except the eccentricity has been increased to 0.15 to show the asymmetric effect on the climate, in this case creating a more persistent polar cap in the southern hemisphere.  The third model is the same as the first, but shows the effect of increasing the obliquity to 75 degrees.

Using this model, we computed approximately 4400 test cases for the
17 models over a range of semi-major axes from 0.80 to 3.0 and
opacities ranging from 0.095, obtained from tuning the Earth baseline
model to 288 K, to 0.26, nearly three times the amount of absorbing
material in the atmosphere.  For each time step in the orbital model,
we run an instance of the energy balance model, computing the three
year climate run and average the last two years to compute the global
mean temperature. In addition, we compute two additional measures of
the habitability of the surface, the Temperature Habitability Index
(THI) and the Ice Habitability Index (IHI).  THI is defined here as
the time-averaged fraction of the surface that is between 273 K to 373
K. IHI is defined as the time-averaged ice-free regions of the planet
regardless of surface temperature. We then normalize the quantities to
the THI and IHI for the $t = 0$ baseline Earth, which has a THI = 0.766 and IHI = 0.485. If the THI (or IHI) is greater than 1.0, this planet is ``more habitable'' than the baseline Earth. If the THI or IHI are less than 1.0, it is less habitable. A THI of 0 indicates a world that is either frozen (in which case the IHI will also be 0) or has no temperatures below 373. The maximum values for THI and ICI by this metric are 1.3 and 2.1, respectively.

This method has some distinct advantages for exploring the effects of the orbit on climate. Since the model is seasonally resolved, even some planets that might have an annually averaged global mean surface temperature below freezing may still experience significant periods of time in the ``habitable'' range and, due to the orbital effects, avoid a planet-wide snowball.

\section{Results}
\label{sec:results}

\begin{figure}
\includegraphics[scale=.40]{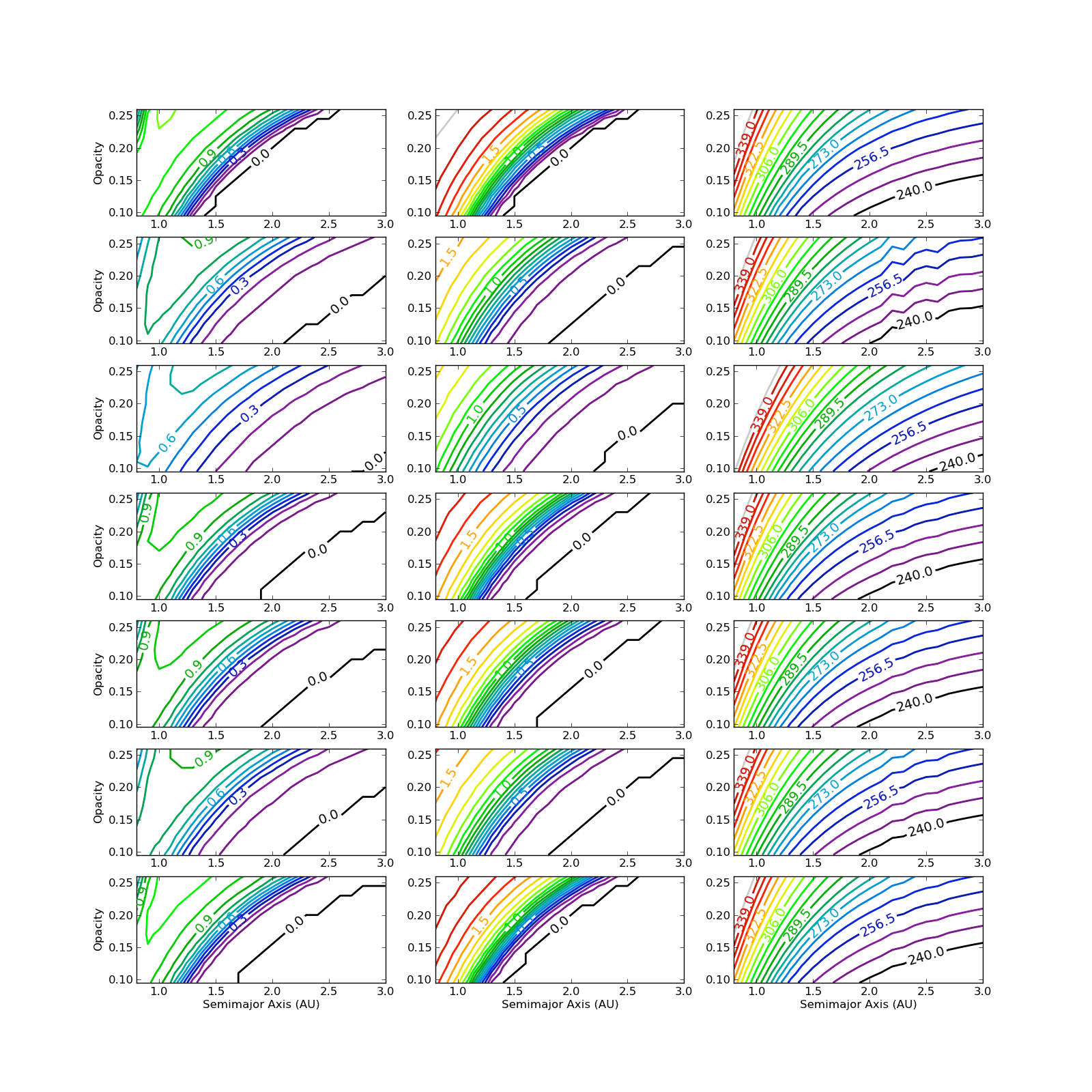}
\caption{The temperature habitability index (THI, left column), the ice habitability index (IHI, middle column), and the mean global temperature (right column) for seven systems listed in Section \ref{sec:obl}.  From top to bottom: System 1 (Baseline) through System 7.}
\label{fig:hab}
\end{figure}

Figure \ref{fig:hab} shows the aggregate results for the seven systems studied in detail. The first column is the THI, the second column is the IHI, and third column is the global mean temperature, all as a function of both the distance from the host star and the opacity of the atmosphere for Systems 1 - 7. Since our simplified EBM ignores complications due to the runaway greenhouse, we will restrict our discussion to the outer edge of the habitable zone. 

The baseline model shows the outer edge of the habitable zone is
approximately 1.4 AU for Earth at $\tau = 0.095$. Each of the other
high obliquity systems and/or high eccentricity systems show a
systematic --- and sometimes dramatic --- increase in the outer edge of the habitable zone. Those increases are tabulated in Table \ref{tab:climatesummary}. The maximum increase is System 3, which has both high eccentricity and large obliquity, with large variations in both parameters.

In order to separate the effects due to the large amplitudes of the eccentricity and obliquity and their variations, we performed static calculations using the mean values of $e$ and $\psi$ to compute the static outer edge. By comparing this value to the outer edge of the habitable zone in the variable runs, $l_{out}$, we can determine how much of the expansion of the outer edge is due to the variability and how much is due to the large values of $e$ and $\psi$.

Table \ref{tab:climatesummary} lists this comparison as the HZ enhancement factor, as a percentage, due to the static orbital properties of the simulations, 

\begin{equation}
	E_S = \left(\frac{l_{static}}{l_{baseline}} - 1\right) \times 100,
\end{equation}

\noindent and the HZ enhancement factor due to the variability of those parameters, 

\begin{equation}
	E_V = \left(\frac{l_{out}}{l_{static}} - 1\right) \times 100,
\end{equation}

\noindent where $l_{baseline}$ is the outer edge of the baseline
system, and $l_{static}$ is the outer edge in the static case. From
this analysis, we see that the increase in the outer edge is dominated
by the large values of $e$ and $\psi$. However, a non-negligible
component --- in one case the sole component, as in System 7 --- is due to the variability of those values.

\begin{figure}
\center
\includegraphics[scale=0.7]{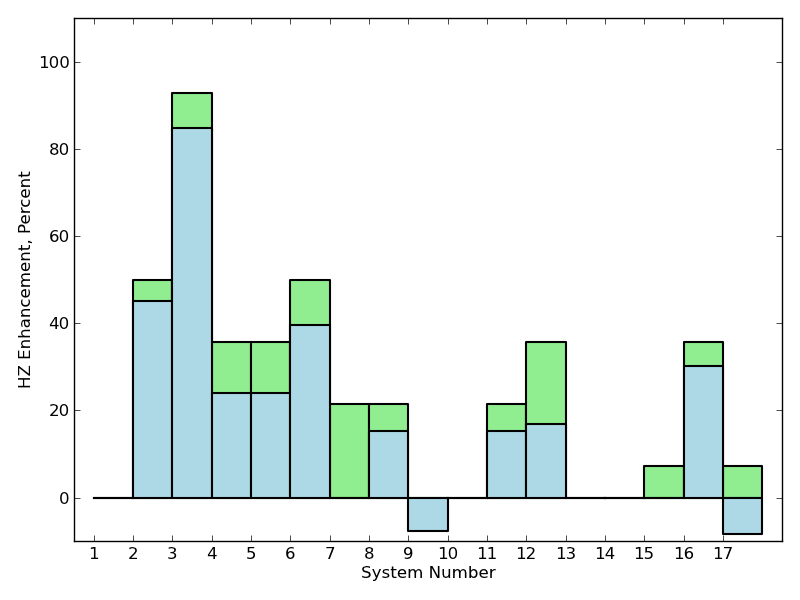}
\caption{A visualization for the HZ enhancement factors from Table \ref{tab:climatesummary}. The height of the bars is the HZ enhancement factor, $E_S$, for the complete simulations. The green box shows the fraction of that enhancement due to the variability of the system, $E_V$. }
\label{fig:eof}
\end{figure}

The values for $E_S$ and $E_V$ are listed in Table \ref{tab:climatesummary} and shown schematically in Figure \ref{fig:eof}. In the figure, the height of the bar is equal to $E_S$ for each system. The green portion of the bar indicates the percentage of the enhancement that is due to the variability alone. For example, our most enhanced system, System 3, had a 93\% increase in the HZ compared to the baseline model.  Of that, 8 \% of the enhancement is due to the variability of the system. In one case, System 7, the enhancement is entirely due to the variability of the parameters.

In some cases, the non-variable systems move the HZ inward from the baseline value of 1.4. System 9 shows no enhancement in the full simulation, but the HZ moves inward when variability is removed (hence the negative ``enhancement''). In System 17, the static run produces an outer limit that is 7 \% smaller than the baseline system (again causing negative enhancement), but dynamics allow a 15 \% increase compared to its static value. 

\begin{figure}
\includegraphics[scale=0.7]{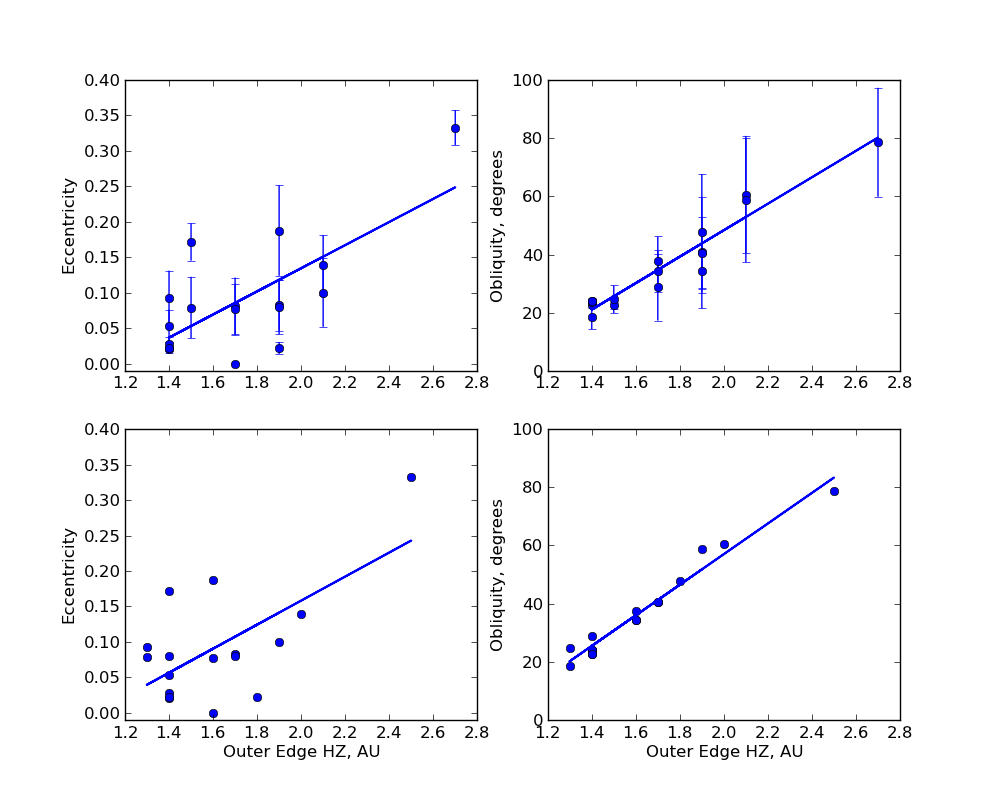}
\caption{A comparison of how the eccentricity and obliquity impact the calculated outer edge of the habitable zone as determined by the temperature habitability index for the baseline opacity of 0.095. The top panels show the eccentricity (left) and obliquity (right) for the variable runs. The bottom panels show the same for the static cases. The error bars for the simulations represent the standard deviation in the eccentricity and the obliquity. Note there is little correlation with eccentricity, but a strong correlation between obliquity and the outer edge of the habitable zone. When the variability in the runs is removed, the outer edge moves inward in all but four cases (see Table \ref{tab:climatesummary}). Values for the linear fits are given in Table \ref{tab:fitparams}.}
\label{fig:habhaip}
\end{figure}

Figure \ref{fig:habhaip} compares the outer edge of the habitable zone for the variable cases (top panels) and static cases (bottom panels) as a function of the eccentricity and obliquity. The error bars on Systems 1 - 17 are derived from the standard deviation of the eccentricity and obliquity from the complete simulations. The points in the top panels lie to right of the those in the bottom showing that the variability increases the habitable zone. However, the effect is most strongly correlated with the obliquity. 

In an effort to quantify the relationship, Table \ref{tab:fitparams} lists the linear regression slope and intercept, along with an error-weighted goodness of fit indicator,

\begin{equation}
	\chi^2 = \sum \frac{\left(y_i - f(x_i)\right)^2}{\sigma_i^2}, 
\end{equation}

\noindent where $y_i$ is the ``observed'' value of either $\psi$ or $e$, $f(x_i)$ is the linear fit, and $\sigma_i$ is the uncertainty computed for the given parameter. In the non-variable cases, since we have no estimate of the uncertainly, the errors in the goodness of fit calculation are taken as 1 \% of the value of the data point. The results show little correlation with eccentricity, but a very strong relationship between obliquity and the outer edge of the HZ. Removing the variability reduces the intercept by 5 degrees and steepens the slope by 7 degrees per AU, which means the outer edge systematically moves inward when the variability is removed.

\section{Discussion}

Our simulations show that the evolution of planets' orbit and rotation can increase the maximum separation between a star and a habitable planet by up to 93\%. By controlling for the natural extension due to larger eccentricity and obliquity, we find that their oscillations can extend the outer edge by up to 20\%, and never decrease it. Thus, the number of potentially habitable planets in the galaxy may be larger than previously thought.

We interpret our results to mean that planets with large and rapid obliquity oscillations are more likely to be habitable than those with negligible oscillations, such as the Earth. This perspective is at odds with the notion that the stability of the Earth's obliquity is important to the development of life. While it still may be true that rapid oscillations can be detrimental, and certainly at some point obliquity cycles could be too large and rapid, our results clearly show that rapid obliquity evolution can be a boon for habitability. At the least, one should not rule out life on planets with rapid obliquity cycles.

Our results are important for future telescopic searches for life, such as the Terrestrial Planet Finder (TPF). Although a final design has yet to be selected, TPF's mission is to directly image potentially habitable planets. These observations depend critically on large star-planet separations in order to disentangle stellar light from reflected planetary light. Our results show that potentially habitable planets can exist at larger star-planet separations than previously appreciated, improving the odds that TPF can discover an inhabited planet.

By necessity, our study was based on hypothetical planets. While our model systems are extreme from a Solar System point of view, they were relatively tame by exoplanet standards. Eccentricities larger than 0.9 have been discovered \citep{2001A&A...375L..27N, 2006MNRAS.369..249J,2008A&A...480L..33T}, very large mutual inclinations are implied by the misalignment between some star's rotation axis and the orbital planet of a companion planet \citep{Triaud10,Naoz11}, and eccentricity-inclination coupling can drive large and rapid oscillations, if both properties are large \citep{Kozai62,2007ApJS..168..297T,Barnes11}. Thus, our results should not be viewed as an extreme possibility, but rather as in the middle of a spectrum of spin-orbit coupling.

Our approach has been simplified in several important ways. While our N-body simulations are the best representation of possible orbits, our rotational model is simplified. A better approach would be to calculate the direct torques from all the bodies in the system and adjust the angular momentum distribution accordingly. Such a model is much more computationally expensive, but could be incorporated into an N-body model without too much extra computational cost. Our EBM is also highly idealized and future improvements could include ocean/land dichotomies, the physics of glacier advancement and retreat, cloud physics, and ultimately even a 3-dimensional global circulation model. Each of these additions, however, adds free parameters to a model with very few constraints. While these features will improve realism, we will continue to suffer from a dearth of observational constraints. Nonetheless, as we move toward identifying planets worthy of detailed spectroscopic followup, such modeling could provide additional insight for prioritization.

While our results demonstrate that planetary system architecture can influence the position of the habitable zone, it remains unclear how
robust this influence is. For example, our planets all began with a spin rate of 24 hours and an obliquity of 23.5$^\circ$. How do different choices change the picture presented here? Future work should explore a range of initial conditions and determine if certain architecture always drive the planet into a particular obliquity cycle. If true, then we may be able to characterize a planet's
obliquity without direct measurements. While such a study was beyond the scope of this paper, the possibility of tightly constraining obliquity is tantalizing and certainly worthy of a follow-up investigation.

Our study suggests that rapid changes in obliquity and eccentricity increase the outer edge of the HZ. We quantify that relationship with linear trends in the enhancement factor with obliquity, but we did not find a threshold to achieve a specific quality that permits significant expansion. We blame the small number of systems we studied for this ambiguity, and leave its identification for future work. We note that prior to running a simulation, it is very difficult to know how the orbital and rotational angular momenta will evolve, thus it could take considerable effort to produce a suite of architectures that suitably cover parameter space.

Our study has shown how orbital architecture is a crucial factor when assessing planetary habitability. While previous work has mostly
focused on static planetary properties, planets are expected to lie in multi-planet systems and hence the sequence of states must be considered. For the foreseeable future, we will have very few constraints on the properties of potentially habitable planets and we therefore must leverage any information we have. 

\label{sec:discussion}

\acknowledgements{This work was supported by the NASA Astrobiology Institute's Virtual Planetary Lab lead team, and an award from the NAI Director's Discretionary Fund.}

\bibliographystyle{apalike}    
\bibliography{/Users/jcarmstrong/Dropbox/Papers/masterbiblio}


\clearpage

\begin{longtable}{llllll} 
\hline
 {\em System } & {\em N} & {\em $\bar{a}$}, AU & {\em $\bar{e}$} & {\em $\bar{i}$}, deg  & {\em M}, $M_{\oplus}$\\ \hline \hline
1 & 1 &  1.0 &  0.05 &  1.4 &  0.3 \\ 
         & 2 &  5.2 &  0.05 &  0.9 &  332.9 \\ 
         & 3 &  9.5 &  0.05 &  1.7 &  133.2 \\ 
2 & 1 &  1.0 &  0.14 &  17.5 &  1.0 \\ 
         & 2 &  0.4 &  0.30 &  19.2 &  10.0 \\ 
         & 3 &  3.0 &  0.30 &  7.9 &  10.0 \\ 
3 & 1 &  1.0 &  0.33 &  20.2 &  1.0 \\ 
         & 2 &  10.0 &  0.08 &  17.7 &  3178.0 \\ 
         & 3 &  19.5 &  0.07 &  12.5 &  3178.0 \\ 
4 & 1 &  1.0 &  0.09 &  11.3 &  1.0 \\ 
         & 2 &  0.1 &  0.30 &  10.9 &  3178.0 \\ 
         & 3 &  34.0 &  0.64 &  2.2 &  3178.0 \\ 
5 & 1 &  1.0 &  0.08 &  30.1 &  1.0 \\ 
         & 2 &  5.0 &  0.30 &  0.0 &  3178.0 \\ 
6 & 1 &  1.0 &  0.10 &  10.0 &  1.0 \\ 
         & 2 &  5.0 &  0.30 &  0.0 &  317.8 \\ 
7 & 1 &  1.0 &  0.08 &  30.0 &  1.0 \\ 
         & 2 &  5.0 &  0.30 &  0.0 &  317.8 \\ 
8 & 1 &  1.0 &  0.0001 &  15.3 &  1.0 \\ 
         & 2 &  0.4 &  0.00002 &  19.6 &  10.0 \\ 
         & 3 &  2.5 &  0.0001 &  8.6 &  10.0 \\ 
9 & 1 &  1.0 &  0.02 &  2.3 &  1.0 \\ 
         & 2 &  14.9 &  0.07 &  11.7 &  3178.0 \\ 
         & 3 &  29.3 &  0.09 &  8.3 &  3178.0 \\ 
10 & 1 &  1.0 &  0.08 &  21.2 &  1.0 \\ 
         & 2 &  0.1 &  0.30 &  20.5 &  3178.0 \\ 
         & 3 &  34.2 &  0.64 &  3.9 &  3178.0 \\ 
11 & 1 &  1.0 &  0.08 &  32.7 &  1.0 \\ 
         & 2 &  0.1 &  0.30 &  30.8 &  3178.0 \\ 
         & 3 &  34.7 &  0.64 &  5.6 &  3178.0 \\ 
12 & 1 &  1.0 &  0.19 &  20.8 &  1.0 \\ 
         & 2 &  0.6 &  0.21 &  21.4 &  1.0 \\ 
         & 3 &  2.5 &  0.30 &  23.3 &  1.0 \\ 
13 & 1 &  1.0 &  0.02 &  7.2 &  1.0 \\ 
         & 2 &  10.0 &  0.06 &  6.4 &  317.8 \\ 
         & 3 &  29.9 &  0.04 &  3.7 &  317.8 \\ 
14 & 1 &  1.0 &  0.02 &  14.3 &  1.0 \\ 
         & 2 &  10.0 &  0.06 &  12.7 &  317.8 \\ 
         & 3 &  29.9 &  0.04 &  7.3 &  317.8 \\ 
15 & 1 &  1.0 &  0.19 &  21.9 &  1.0 \\ 
         & 2 &  10.0 &  0.06 &  19.1 &  317.8 \\ 
         & 3 &  29.9 &  0.04 &  10.9 &  317.8 \\ 
16 & 1 &  1.0 &  0.02 &  14.0 &  1.0 \\ 
         & 2 &  5.0 &  0.04 &  10.8 &  317.8 \\ 
         & 3 &  10.0 &  0.06 &  19.3 &  127.1 \\ 
17 & 1 &  1.0 &  0.08 &  30.0 &  1.0 \\ 
         & 2 &  5.0 &  0.30 &  0.1 &  127.1 \\ \hline \hline
\caption{\narrower Summary of 17 systems modeled including the mean semi major axis $\bar{a}$, mean eccentricity, $\bar{e}$, mean inclination, $\bar{i}$, and planet mass, $M_p$, in Earth masses. The systems consist of either two or three planets, one being an Earth-mass planet located at 1.0 AU.} 
\label{tab:systemsummary}
\end{longtable}

\clearpage

\begin{table}[htp]
\centering
\begin{tabular}{llll} \hline
 {\em Parameter } & {\em Symbol}  & {\em Value} & {\em Units }  \\ \hline \hline
 Gravity constant & $k$    &   0.01720209895 &   $AU^3 \ M_{\odot}^{-1} \ day^{-2}$ \\
 Angular spin rate & $\nu$    &   $2 \pi$ &    $rad \ day^{-1}$\\
 Dynamic ellipticity & $E_D$   &   0.00328005 &  $unitless$     \\
 Initial obliquity & $\psi_0$ &    23.44 & $degrees$    \\
 Initial precession & $p_A$ &    0.0 & $degrees$    \\
 Planet mass &  $M_p$ &   1.0   & $Earth \ masses$ \\
\hline \hline
\end{tabular}
\caption{\narrower Parameters for the obliquity calculation.} 
\label{tab:oblparams}
\end{table}

\clearpage

\begin{table}[htp]
\centering
\begin{tabular}{lllll} \hline
 {\em Parameter } &   {\em Symbol }    & {\em Value} & {\em Units }  \\ \hline \hline
 Stellar mass   & $M_{*}$ & 1.0 & $Solar \ masses$ \\
 Stellar luminosity & $L_{*}$ & 1.0 & $Solar \ luminosities$ \\
 Land albedo &   $A_{land}$    &   0.4 &   $unitless$ \\
 Ice albedo &   $A_{ice}$      &   0.6 &   $unitless$ \\
 Surface heat flux &  $F_{surf}$   & 88 & $watts$ \\
 Baseline opacity & $\tau $ & 0.095 & $unitless$ \\
 Surface emissivity &   $\epsilon_{surf}$ & 1.0 & $unitless$ \\
 Ice/snow deposition rate & $R_{ice}$ & $5.0\times10^{-5}$ & $kg \ m^{-2} \ s^{-1}$ \\ 
\hline \hline
\end{tabular}
\caption{\narrower Parameters for the climate calculation. The parameters of the climate model are selected to be within reasonable limits and still reproduce climates similar to Earth under current conditions. The only other free parameters are the atmospheric opacity and the distance from the star, both varied as part of the study.} 
\label{tab:clmparams}
\end{table}

\clearpage

\begin{table}[htp]
\centering
\begin{tabular}{lllllllll} \hline
{\em N } & {\em $\bar{e}$} & {\em $P_e$}, Myr & {\em $\bar{\psi}$}, deg & {\em $P_{\psi}$}, Myr & {\em $l_{out}$}, AU &  {\em $l_{static}$}, AU &{\em $E_S$, \%} & {\em $E_T$, \%} \\ \hline \hline
1 & 0.05$\pm$0.023 & 0.25 & 22.6$\pm$1.1 & 0.17 & 1.4&1.4 &  n/a&  n/a \\ 
2 & 0.14$\pm$0.042 & 0.20 & 60.6$\pm$20.2 & 0.33 & 2.1&2.0 & 50 &  5 \\ 
3 & 0.33$\pm$0.025 & 0.04 & 78.6$\pm$18.7 & 0.17 & 2.7&2.5 & 93 &  8 \\ 
4 & 0.08$\pm$0.036 & 0.02 & 40.7$\pm$12.1 & 0.02 & 1.9&1.7 & 36 & 12 \\ 
5 & 0.08$\pm$0.038 & 0.05 & 40.6$\pm$19.1 & 0.02 & 1.9&1.7 & 36 & 12 \\ 
6 & 0.10$\pm$0.048 & 0.14 & 58.7$\pm$21.4 & 0.33 & 2.1&1.9 & 50 & 11 \\ 
7 & 0.08$\pm$0.041 & 0.07 & 28.8$\pm$11.4 & 0.17 & 1.7&1.4 & 21 & 21 \\ 
8 & 0.0001$\pm$0.00006 & 0.007 & 37.6$\pm$8.6 & 0.23 & 1.7&1.6 & 21 &  6 \\ 
9 & 0.09$\pm$0.038 & 0.01 & 18.4$\pm$4.1 & 0.02 & 1.4&1.3 &  -7 &  8 \\ 
10 & 0.02$\pm$0.001 & 0.08 & 23.9$\pm$0.7 & 0.04 & 1.4&1.4 &  0 &  0 \\ 
11 & 0.08$\pm$0.035 & 0.01 & 34.3$\pm$7.1 & 0.02 & 1.7&1.6 & 21 &  6 \\ 
12 & 0.19$\pm$0.064 & 0.33 & 34.3$\pm$7.4 & 0.20 & 1.9&1.6 & 36 & 19 \\ 
13 & 0.03$\pm$0.010 & 0.33 & 23.7$\pm$0.6 & 0.09 & 1.4&1.4 &  0 &  0 \\ 
14 & 0.02$\pm$0.008 & 0.25 & 24.2$\pm$1.0 & 0.09 & 1.4&1.4 &  0 &  0 \\ 
15 & 0.17$\pm$0.026 & 0.25 & 22.6$\pm$1.2 & 0.08 & 1.5&1.4 &  7 &  7 \\ 
16 & 0.02$\pm$0.008 & 0.33 & 47.9$\pm$19.8 & 0.20 & 1.9&1.8 & 36 &  6 \\ 
17 & 0.08$\pm$0.043 & 0.17 & 24.6$\pm$4.7 & 0.10 & 1.5&1.3 &  -7 & 15 \\ \hline \hline
\end{tabular}
\caption{\narrower Summary of the systems used in the complete climate comparison, including the system number, $N$, the mean eccentricity, $\bar{e}$, along with its dominate period of oscillation, $P_e$ in Myr, mean obliquity, $\bar{\psi}$, with its period, $P_{\psi}$, the calculated outer edge of the habitable zone, $l_{out}$, for temperature based on the habitability index, and the orbit enhancement factor ($E_S$) and the variable enhancement factor ($E_V$), see text.} 
\label{tab:climatesummary}
\end{table}

\clearpage

\begin{table}[htp]
\centering
\begin{tabular}{llll} \hline 
{\em  } & {\em Slope} & {\em Intercept}  & {\em $\chi^{2}$}  \\ \hline \hline
Variable Eccentricity & 0.16 & -0.19 & $1.7\times10^5$ \\
Average Eccentricity & 0.17 & -0.18 & $3.5\times10^8$ \\
Variable Obliquity & 46 & -43 & 4.1 \\
Average Obliquity & 53 & -48 & 91 \\ \hline \hline
\end{tabular}
\caption{\narrower Parameter fits for the linear models in Figure \ref{fig:habhaip}. The calculation of the goodness of fit parameter, $\chi^{2}$, along with the associated uncertainties, is described in the text.} 
\label{tab:fitparams}
\end{table}



\end{document}